\documentclass[12pt]{article}
\usepackage{amssymb}
\usepackage{epsfig}

\catcode`\@=11
\def\numberbysection{\@addtoreset{equation}{section}
        \def\theequation{\thesection.\arabic{equation}}}

\def\beq{\begin{equation}}
\def\eeq{\end{equation}}
\numberbysection
\begin{document}
\begin{titlepage}
\begin{center}
\hfill DFF  2/1/05 \\
\vskip 1.in {\Large \bf ($2+1$) noncommutative gravity and conical
spacetimes} \vskip 0.5in P. Valtancoli
\\[.2in]
{\em Dipartimento di Fisica, Polo Scientifico Universit\'a di Firenze \\
and INFN, Sezione di Firenze (Italy)\\
Via G. Sansone 1, 50019 Sesto Fiorentino, Italy}
\end{center}
\vskip .5in
\begin{abstract}
We solve ($2+1$) noncommutative gravity coupled to point-like
sources. We find continuity with Einstein gravity since we recover
the classical gravitational field in the $\theta \rightarrow 0$
limit or at large distance from the source. It appears a
limitation on the mass which is twice than expected. Since the
distance is not gauge invariant, the measure of the deficit angle
near the source is intrinsically ambiguous, with the gauge group
playing the role of statistical ensemble. Einstein determinism can
be recovered only at large distance from the source, compared with
the scale of the noncommutative parameter $\sqrt{\theta}$.
\end{abstract}
\medskip
\end{titlepage}
\pagenumbering{arabic}
\section{Introduction}

General relativity has always been regarded as one of the most
important achievements of mankind, despite some philosophers have
criticized it for its strict determinism.

In the meantime the level of abstraction in mathematics has
increased from Einstein's age and it has permitted us conceiving
spaces in which the locality principle is lost. Nowadays
space-time can be substituted by a noncommutative algebra and the
ordinary differential structures, i.e. derivative and integration,
can be generalized to more abstract operations such as the
commutator and the trace. Non-commutativity, connected to these
structures, has certain analogy with quantum mechanics, so one
might expect that, once inserted in the physical laws, it produces
a new fundamental limit on the possibility of determining and
measuring the physical quantities, which are object of
experimental tests.

Following such general reasoning, we have decided to study in a
concrete example how the inspiring principles of general relativity
are extended in the noncommutative case \cite{3}-\cite{11}-\cite{5}.
We have chosen the case of ($2+1$) gravity \cite{2}-\cite{10}
coupled to a point mass source \cite{9}-\cite{8}, where meaningful
results are simple to reach.

In classical Einstein gravity mass is the primary source for
deforming space-time, and a test particle is gravitationally
influenced by a mass source because it follows the geodesic of a
deformed space-time.

In the literature ( \cite{9} ) it is questioned that mass is the
source for deforming space-time in noncommutative gravity. Our
first objective was then to build an explicit solution of the
noncommutative equations of motion confirming that mass is really
the only cause for deforming space-time. To achieve such result we
have taken advantage of the definition given in \cite{9} of
point-like source, an extended source which only in the $\theta
\rightarrow 0$ limit reduces to a delta-function singularity. We
have then rewritten the equations of motion in the form of a
commutator between operators, following the matrix model
formalism. With simple manipulations we can prove the existence of
a non trivial solution for the spin connection and the vierbein.

By carrying out the correct classical limit, we can show that the
noncommutative gravitational field is a smooth deformation of the
classical one \cite{6}-\cite{7}. Moreover it is possible to extend
such results to the massive and spinning case, i.e. in presence of a
torsion source.

At this point it is natural to rise the following question. In
what sense noncommutative gravity can be distinguished from
general relativity ? What is its characteristic signal ?

To answer such question we have to verify the other inspiring
principle of general relativity, i.e. that the influence on a test
particle of a massive source is determined by the geodesic motion
on a deformed space-time.

In ($2+1$) dimensions the scattering angle of a test particle is
strictly related to the deficit angle of the conical space-time.
To reveal such deficit angle it is necessary integrating the
distance on a circumference centered on the mass source. However
here we encounter a serious problem, since the distance is not a
gauge invariant concept in noncommutative gravity, differently
from the classical case. Two options are open: i) finding a
gauge-invariant quantity substituting the distance in the
noncommutative case ( hard task ) , ii)  working with the
distance, as it is defined in general relativity, keeping in mind
that its meaning is intrinsically ambiguous since it is not gauge
invariant.

In the final discussion of this paper we attempt to give a partial
answer to this question, reducing such arbitrariness in the
proximity of the source, by imposing that the internal gauge
transformations reduce to a constant transformation at spatial
infinity. Proceeding this way we can at least assure continuity
with general relativity. At large distance from the source,
compared with the scale $\sqrt{\theta}$ of the noncommutative
parameter, both theories give coincident predictions, but at small
distance from the source Einstein determinism is irremediably
lost, and the reason of such indeterminism comes from the clash
between the concept of distance and the noncommutative symmetries
of the model.

\section{Lagrangian of ($2+1$) noncommutative gravity}

Noncommutative gravity is a modification of classical Einstein
gravity compatible with a noncommutative algebra between the
coordinates. In ($2+1$) dimensions a particularly convenient example
for such an algebra is represented by a noncommutative plane:

\beq [ \hat{x} , \hat{y} ] = i \theta \ \ \ \Rightarrow \ \ \ \ \
[ \hat{z} , \hat{\overline{z}} ] = 2 \theta \ \ \ \ \ \ \hat{z} =
\hat{x} + i \hat{y} \label{21}\eeq

The noncommutative analogue of the Einstein action can be easily
obtained in the first order formalism by gauging the internal
$SO(2,1)$ Lorentz invariance, extended to $U(1,1)$ gauge group.
The generators of $U(1,1)$ group are defined as

\beq \tau_1 = i \sigma_1  \ \ \ \ \ \tau_2 = i \sigma_2 \ \ \ \ \
\tau_3 = \sigma_3 \label{22}\eeq

They satisfy the following hermiticity conditions

\beq \tau^{\dagger}_i = \tau_3 \tau_i \tau_3 \label{23}\eeq

and the composition property

\beq \tau_i \tau_j = \eta_{ij} - i \epsilon_{ijk} \eta^{km} \tau_m
\label{24}\eeq

The $U(1,1)$ gauge group is necessary to take into account the
signature of the flat metric tensor $\eta^{\mu\nu} = ( - - + )$,
typical of ($2+1$) dimensions.

We prefer working in the operator formalism instead of the star
product formalism, and in analogy with the vierbein and spin
connection fields we define two basic operators:

\begin{eqnarray}
Y_\mu & = & e_\mu^a \tau_a + e^0_\mu \nonumber \\
X_\mu & = & \hat{p}_\mu + \frac{\omega^a_\mu}{2} \tau_a +
\omega^0_\mu \label{25}\end{eqnarray}

where all the components ($ e^a_\mu , e^0_\mu $) and ($ \omega^a_\mu
, \omega^0_\mu $) are operators acting on the Hilbert space on which
the basic commutation relations (\ref{21}) are represented.

The classical property that the vierbein and spin connection are
real fields can be taken into account in this scheme by requiring
the following hermiticity conditions for the basic matrices
($X_\mu , Y_\mu $) :

\begin{eqnarray}
Y^{\dagger}_\mu & = & \tau_3 Y_\mu \tau_3 \ \ \ \ \ \Rightarrow \
\ \ \ \ \ e^{a \dagger}_\mu = e^a_\mu \ \ \ \ \ e^{0 \dagger}_\mu
= e^o_\mu \nonumber \\
X^{\dagger}_\mu & = & \tau_3 X_\mu \tau_3 \ \ \ \ \ \Rightarrow \
\ \ \ \ \ \omega^{a \dagger}_\mu = \omega^a_\mu \ \ \ \ \
\omega^{0 \dagger}_\mu = \omega^o_\mu \label{26}\end{eqnarray}

All these component are by construction hermitian operators. The
background of the matrix model $\hat{p}_\mu$ has to satisfy the
commutation relations :

\beq [ \hat{p}_\mu , \hat{p}_\nu ] = i \theta^{-1}_{\mu\nu}
\label{27}\eeq

where the tensor on the right hand side is defined for the
noncommutative plane as

\beq \theta_{\mu\nu}^{-1} = \frac{1}{\theta} \left(
\begin{array}{ccc} 0 & 1 & 0 \\ -1 & 0 & 0 \\ 0 & 0 & 0
\end{array} \right) \label{28}\eeq

The classical pure Einstein action on a noncommutative plane is

\beq S_E = \beta \ \epsilon^{\mu\nu\rho} \ Tr \ [ \ Y_\mu \ ( \ [
X_\nu , X_\rho ] - i \theta_{\nu\rho}^{-1} \ ) \ ] \label{29}\eeq

To discuss the properties of this action, it is convenient to
define the curvature and torsion tensors:

\begin{eqnarray}
R_{\mu\nu} & = & [ X_\mu , X_\nu ] - i \theta_{\mu\nu}^{-1}
\nonumber \\
T_{\mu\nu} & = & [ X_\mu , Y_\nu ] - [ X_\nu , Y_\mu ]
\label{210}\end{eqnarray}

By using the results of ref. \cite{2} we can state that the Einstein
action $S_E$ is endowed with a deformed diffeomorphism invariance
defined as

\begin{eqnarray}
\delta_v X_\mu & = & \frac{1}{2} \{ v^\alpha , R_{\mu\alpha} \}
\nonumber \\
\delta_v Y_\mu & = & \frac{1}{2} \{ v^\alpha , T_{\mu\alpha} \}
\label{211}\end{eqnarray}

By substituting these transformations into the action we find

\begin{eqnarray}
\delta_v S_E & = & \frac{1}{2} \epsilon^{\mu\nu\alpha} Tr ( \
\delta_v
Y_\mu R_{\nu\alpha} + \delta_v X_\mu T_{\nu\alpha} \ ) = \nonumber \\
& = & \frac{1}{4} Tr [ \ v^\beta \epsilon^{\mu\nu\alpha} -
v^\alpha \epsilon^{\mu\nu\beta} - v^\mu \epsilon^{\beta\nu\alpha}
- v^\nu \epsilon^{\alpha\mu\beta} \ ] T_{\mu\beta} R_{\nu\alpha}
\label{212}\end{eqnarray}

but the tensor in parenthesis is trivially null. A similar trick
applies in two dimensions \cite{1}-\cite{4}.

These observations are generalizable to the more general action of
($2+1$) gravity containing the cosmological constant term:

\begin{eqnarray}  S_T & = & S_{+} - S_{-} \nonumber \\
S_{\pm} & = & \epsilon^{\mu\nu\alpha} Tr \left[ X^{\pm}_\mu \left(
\frac{1}{3} [ X^\pm_\nu , X^\pm_\alpha ] - i
\theta^{-1}_{\nu\alpha} \right) \right] \nonumber \\
X^{\pm}_\mu & = & X_\mu \pm Y_\mu \label{213}\end{eqnarray}

In this case the generator of extended general covariance is
defined as

\begin{eqnarray}
\delta_v X^{\pm}_\mu & = & \frac{1}{2} \{ v^\alpha ,
R^\pm_{\alpha\mu} \} \nonumber \\
R^\pm_{\mu\nu} & = & [ X^\pm_\mu , X^\pm_\nu ] - i
\theta^{-1}_{\mu\nu} \label{214}\end{eqnarray}

The equations of motion of ($2+1$) noncommutative Einstein gravity
coincide with the vanishing of both curvature and torsion tensors:

\begin{eqnarray}
R_{\mu\nu} & = & [ X_\mu , X_\nu ] - i \theta_{\mu\nu}^{-1} = 0
\nonumber \\
T_{\mu\nu} & = & [ X_\mu , Y_\nu ] - [ X_\nu , Y_\mu ] = 0
\label{215}\end{eqnarray}

These conditions are mapped into themselves by gauge invariance

\begin{eqnarray}
X_\mu & \rightarrow & U^{-1} X_\mu U \nonumber \\
Y_\mu & \rightarrow & U^{-1} Y_\mu U \label{216}\end{eqnarray}

where $U$ is a $U(1,1)$ gauge transformation satisfying the
unitarity condition

\beq U^\dagger \tau_3 U \tau_3 =  1 \label{217}\eeq

and by general covariance.

The classical limit ( $\theta \rightarrow 0$ limit ) is obtained
by imposing that the background $\hat{p}_\mu$ has the following
representation on the symbols of the operators ($ e^a_\mu, ... ,
\omega^a_\mu, .... $):

\beq [ \hat{p}_\mu , ... ] \rightarrow -i \partial_\mu ( ... )
\label{218}\eeq

While the classical limit is better understood in the star product
formalism, we will prefer to solve the gravitational equations of
motion directly in the operator formalism, where we will find a
considerable simplification allowing us to generalize the typical
conical space-times of Einstein gravity.

\section{Conical solutions: spin connection}

Our motivation is to study the deformation of space-time induced by
a static mass source, in presence of a noncommutative plane. The
first step is the definition of a mass source, which has been
already solved in ref. \cite{9}, with the introduction of
"point-like" source ( although we are afraid that this coupling
doesn't respect any kind of general covariance for  $\theta \neq 0$
). This is a distributed source which only in the $\theta
\rightarrow 0$ limit produces a singular delta-function source. In
the Hilbert space there is in fact a natural candidate for such a
source, i.e. a simple operator whose symbol is a well known
representation of a delta-function:

\beq {\rm Pointlike \ source } \ \ \equiv P_0 = | 0 >< 0 |
\label{31}\eeq

The symbol corresponding to $P_0$ is in fact

\beq P_0 = | 0 >< 0 | \longrightarrow 2 e^{-\frac{r^2}{\theta}}
\label{32}\eeq

and the representation of a delta-function can be realized as:

\beq \lim_{\theta \rightarrow 0} \ \frac{1}{2\pi\theta} \ P_0 =
\delta^2 (x) \label{33}\eeq

Formula (\ref{32}) is a particular example of the general
transformation rule between operators and symbols on the
noncommutative plane:

\beq | n >< n+l | \equiv 2 (-1)^n \sqrt{ \frac{ n ! }{ (n+l) ! }}
{\left( \frac{2 r^2 }{\theta} \right)}^{\frac{l}{2}} L^l_n {\left(
\frac{2 r^2 }{\theta} \right)}^{\frac{l}{2}}
e^{-\frac{r^2}{\theta}} e^{i l \phi} \label{34}\eeq

where $L^l_n (z)$ are the generalized Laguerre polynomials. For
example the coordinate $z$ can be replaced by the following
operator:

\begin{eqnarray}
\hat{z} & = & \sqrt{ 2\theta } \sum^{\infty}_{n=0} \sqrt{ n+1 } |
n >< n+1 | \nonumber \\
\hat{\overline{z}} & = & \sqrt{ 2\theta } \sum^{\infty}_{n=0}
\sqrt{ n+1 } | n+1 >< n | \label{35}\end{eqnarray}

and it automatically satisfies the commutation relations:

\beq [ \hat{z} , \hat{\overline{z}} ] = 2 \theta \label{36}\eeq

In the classical solutions of ($2+1$) gravity \cite{6}-\cite{7} the
inverses of $z$ and $\overline{z}$ appear. To match the
noncommutative solution with the standard results it is useful to
introduce the following operators:

\begin{eqnarray}
\frac{1}{\hat{z}} & = &
\frac{1}{\sqrt{2\theta}}\sum^{\infty}_{n=0} \frac{1}{\sqrt{ n+1 }}
| n+1 >< n | = \frac{1}{r} ( 1 - e^{-\frac{r^2}{\theta}} ) e^{-i
\phi } \nonumber \\
\frac{1}{\hat{\overline{z}}} & = &
\frac{1}{\sqrt{2\theta}}\sum^{\infty}_{n=0} \frac{1}{\sqrt{ n+1 }}
| n >< n+1 | = \frac{1}{r} ( 1 - e^{-\frac{r^2}{\theta}} ) e^{i
\phi } \nonumber \\
\frac{1}{\hat{z}} \hat{z} & = & \hat{\overline{z}}
\frac{1}{\hat{\overline{z}}} = 1 - | 0 >< 0 |
\label{37}\end{eqnarray}

Therefore the classical formula

\beq \partial_{\overline{z}} \frac{1}{z} = \pi \delta^2 (x)
\label{38}\eeq

can be replaced in the noncommutative theory by:

\beq \frac{1}{2\theta} \left[ \hat{z} , \frac{1}{\hat{z}} \right]
= \frac{1}{2\theta} | 0 >< 0 | = \frac{1}{\theta} \
e^{-\frac{r^2}{\theta}} \ \stackrel{\mbox{\small {$\theta
\rightarrow 0$} }}{\longrightarrow} \ \pi \delta^2(x)
\label{39}\eeq

We are ready to discuss the solution for the spin connection in
presence of a massive source. Let us define the following complex
combinations of $X_\mu$ matrices:

\begin{eqnarray}
X & = & X_z = \frac{1}{2} ( X_1 - i X_2 ) = \hat{p}_z +
\frac{\omega_z^a}{2}
\tau_a + \omega_z^0 \nonumber \\
\overline{X} & = & \overline{X}_{\overline{z}} =
\hat{p}_{\overline{z}} + \frac{\omega_{\overline{z}}^a}{2}
\tau_a^\dagger + \omega_{\overline{z}}^0 \nonumber \\
\hat{p}_z & = & \frac{1}{2} ( \hat{p}_1 - i \hat{p}_2 )
\rightarrow [ \hat{p}_z , ... ] \ \stackrel{\mbox{\small {$\theta
\rightarrow 0$} }}{\longrightarrow} \ - i
\partial_z (..)
\label{310}\end{eqnarray}

The second step is showing that the equations of motion with
point-like source can be written in the form of a commutator:

\begin{eqnarray}
& \ & [ \ X , \overline{X} \ ] = - \frac{1}{2\theta} \left( 1 -
\frac{m}{4\pi} P_0 \tau_3 \right) \nonumber \\
& \ & \lim_{\theta \rightarrow 0} \ \frac{1}{\theta} \ P_0 = 2 \pi
\delta^2 (x) \label{311}\end{eqnarray}

where $m$ is the mass of the point-like source.

It is enough to show that this equation has as classical limit the
Einstein equation coupled to point sources ( see ref.
\cite{6}-\cite{7} ):

\beq {\left( \partial_{[ \mu} \omega_{\nu ]} + \omega_{[\mu} ,
\omega_{\nu ]} \right)}^a_b = - \epsilon_{\mu\nu\lambda}
\epsilon^a_{bc} v^{\lambda} P^c \delta^2 ( x^{\mu} - \xi^{\mu}
(t)) \ \ \ ( 8 \pi G = 1 \ { \rm units } ) \label{312}\eeq

where the impulse $P^a$ and the velocity vector $v^a$ are defined
as

\beq P^a = m v^a = m \gamma (  \overrightarrow{v}, 1 )
\label{313}\eeq

and $\xi^{\mu} (t)$ is the position vector of the source.

In the simplified case we are discussing of a rest particle ( $
\overrightarrow{v} = 0 $), and introducing the auxiliary spin
connection

\beq \omega^a_\mu = -\frac{1}{2} \eta^{a a'} \epsilon_{a' bc}
\omega_{\mu ,}^{bc} \Leftrightarrow \omega_{\mu,ab} = -
\epsilon_{abc} \omega_{\mu}^c \label{314}\eeq

the equations of motion (\ref{312}) become

\beq \partial_{[ \mu} \omega_{\nu ]}^a + \frac{1}{2} \eta^{a a'}
\epsilon_{a' bc} \omega_{[ \mu}^b , \omega_{\nu ]}^c =
\epsilon_{\mu\nu 3} m \delta^a_3 \delta^{2} (x) \label{315}\eeq

To adjust the classical limit of the commutator equation (\ref{311})
we must choose the background $\hat{p}_\mu$ as

\begin{eqnarray}
\hat{p}_x & = & \frac{1}{\theta} \hat{y} \ \ \ \ \ [ \hat{p}_x ,
.. ] \rightarrow - i \partial_x \nonumber \\
\hat{p}_y & = & - \frac{1}{\theta} \hat{x} \ \ \ \ \ [ \hat{p}_y ,
.. ] \rightarrow - i \partial_y \nonumber \\
\hat{p}_z & = & \frac{i}{2\theta} \hat{\overline{z}}
\label{316}\end{eqnarray}

Then the commutator term can be developed in the $\theta
\rightarrow 0$ limit:

\begin{eqnarray}
[ X_x, X_y ] \ \ & \stackrel{\mbox{\small {$\theta \rightarrow 0$}
}}{\longrightarrow} &  \frac{i}{\theta} - \frac{i}{2} [ (
\partial_x \omega_y^a - \partial_y \omega^a_x ) + \frac{1}{2} \eta^{a a'}
\epsilon_{a' bc} \{ \omega^b_x ,
\omega^c_y \} ] \tau_a \nonumber \\
& = & \frac{i}{\theta} - i \frac{m}{2} \tau_3 \delta^2(x) \ \
\stackrel{\mbox{\small {$\theta \rightarrow 0$} }}{\longleftarrow}
\ \  \frac{i}{\theta} \ \left[ 1 - \frac{m}{4\pi} \ P_0 \tau_3
\right] \label{317}\end{eqnarray}

The second equality can be obtained using the classical equation of
motion (\ref{315}) or as a classical limit of the source of the
commutator equation (\ref{311}). We have then proved the equivalence
of the two equations of motion (\ref{311}) and (\ref{312}).

It is interesting to note that the factor in front of the
projector $P_0$ is proportional to the combination $\mu =
\frac{m}{2\pi}$ entering in the condition ( $ \mu < 1 $ ) for the
existence of conical solutions.

We can anticipate that we will be able to obtain a noncommutative
limit on mass, which is however twice than expected, by imposing
that the solution of the commutator equation respects the
hermicity condition:

\begin{eqnarray}
& \ & X^{\dagger}_\mu = \tau_3 X_\mu \tau_3 \nonumber \\
& \ & [ \ X_x, X_y \ ] \ = \ \frac{i}{\theta} \left[ \ 1 -
\frac{\mu}{2} P_0 \tau_3 \ \right] \ \ \ \ \ \ \Leftrightarrow \ \
\ \ \  \frac{\mu}{2} < 1 \label{318}\end{eqnarray}

Fortunately the solution we are going to discuss is contained in
the following simple ansatz:

\begin{eqnarray}
X & = & i \sum^{\infty}_{n=0} \ ( f(n) - \tau_3 g(n)) \ | n+1 >< n
|
\nonumber \\
\overline{X} & = & - i \sum^{\infty}_{n=0} \ ( f(n) - \tau_3 g(n))
\ | n >< n+1 | \label{319}\end{eqnarray}

The hermicity condition (\ref{26}) is satisfied by the reality
condition on the unknown functions $f(n)$ and $g(n)$:

\beq f^{\dagger}(n) = f(n) \ \ \ \ g^{\dagger} (n) = g(n)
\label{320}\eeq

In the massless limit ($\mu \rightarrow 0$), the commutator
equation is solved by the background, i.e. by the choice

\beq f(n) = \sqrt{ \frac{ n+1 }{ 2\theta } } \ \ \ \ \ \ g(n) = 0
\label{321}\eeq

The commutator equation (\ref{311}) , together with the ansatz
(\ref{319}), produces the following recursive relations:

\begin{eqnarray}
& \ & f^2(n) + g^2(n) = f^2(n-1) + g^2(n-1) + \frac{1}{2\theta} =
\frac{n+1}{2\theta} \nonumber \\
& \ & f(n) g(n) = f(n-1) g(n-1) = \frac{\mu}{8\theta}
\label{322}\end{eqnarray}

which are solved by:

\begin{eqnarray}
& \ & f(n) = \frac{1}{2{\sqrt{2\theta}}} \left( \sqrt{
n+1+\frac{\mu}{2} } +
\sqrt{ n+1-\frac{\mu}{2} } \right) \nonumber \\
& \ & g(n) = \frac{1}{2{\sqrt{2\theta}}} \left( \sqrt{
n+1+\frac{\mu}{2} } - \sqrt{ n+1-\frac{\mu}{2} } \right)
\label{323}\end{eqnarray}

In the $\mu \rightarrow 0$ limit we recover the background solution
(\ref{321}). It is now clear that the hermicity condition is
respected if and only if

\beq  f^{\dagger} (n) = f(n) \ \ g^{\dagger} (n) = g(n) \ \ \
\Leftrightarrow \ \ \ n+1 - \frac{\mu}{2} > 0 \ \ \forall n  \ \ \
\Leftrightarrow \ \ \ \frac{\mu}{2} < 1 \label{324}\eeq

We have found that the operator formalism of noncommutative
gravity has some similarity with the results of classical Einstein
gravity. A better confirmation comes from the discussion of the
classical limit ( see section $5$ ).

\section{Conical solutions: vierbein}

A gravity theory is based on the concept of metric. Therefore to
complete the solution for point-like sources we need to extract
the vierbein from the null torsion condition. In the massless case
the natural choice is the flat vierbein

\beq Y_\mu = \delta^a_\mu \tau_a \label{41}\eeq

which can be recast in a more convenient form:

\begin{eqnarray}
Y_\mu & = & i [ X_\mu , \Lambda ] \nonumber \\
\Lambda & = & \hat{x} \tau_1 + \hat{y} \tau_2 + t \tau_3
\label{42}\end{eqnarray}

The null torsion condition

\beq T_{\mu\nu} = [ X_\mu, Y_\nu ] - [ X_\nu, Y_\mu ] = 0
\label{43}\eeq

is automatically satisfied by the ansatz (\ref{42}), since due to
the Jacobi identity

\beq T_{\mu\nu} = i [ [ X_\mu , X_\nu ], \Lambda ] = -
\theta_{\mu\nu}^{-1} [ 1, \Lambda ] = 0 \label{44}\eeq

the equation of motion reduces to a commutator with a c-number,
which is trivially null, for every choice of $\Lambda$. However
requiring that the metric is flat in absence of sources fixes for
$\Lambda$ the form given in eq. (\ref{42}).

Let us study the case $\mu \neq 0$, with $X_\mu$ given by eqs.
(\ref{319}) and (\ref{323}). The ansatz (\ref{42}) is again useful:

\beq Y_\mu = i [ X_\mu, \Lambda ] \ \ \ \ \ \Lambda^{\dagger} =
\tau_3 \Lambda \tau_3 \label{45}\eeq

with $\Lambda$ an unknown operator.

By using the Jacobi identity, since the commutator of two spin
connections is proportional to the projector operator $P_0$, the
null torsion condition is solved by the following condition on
$\Lambda$:

\beq [ \Lambda , P_0 \tau_3 ] = 0 \ \ \ \ \ \Lambda \ \
\stackrel{\mbox{\small {$\theta \rightarrow 0$}
}}{\longrightarrow} \ \ \hat{x}_a \tau_a \label{46}\eeq

A natural choice is dressing the flat solution with quasi-unitary
operators:

\begin{eqnarray}
\Lambda & = & U^{\dagger} \hat{x}_a U \tau_a \nonumber \\
U & = & \sum^{\infty}_{n=0} | n >< n+1 | \label{47}\end{eqnarray}

due to the properties

\beq U P_0 = P_0 U^{\dagger} = 0 \label{48}\eeq

Let us define some new coordinate operators:

\begin{eqnarray}
\hat{z'} & = & \sqrt{2\theta} \sum^{\infty}_{n=0} \sqrt{n} \ | n
><
n+1 | = U^{\dagger} \hat{z} U \nonumber \\
\hat{\overline{z'}} & = & \sqrt{2\theta} \sum^{\infty}_{n=0}
\sqrt{n} \ | n+1
>< n | = U^{\dagger} \hat{\overline{z}} U
\label{49}\end{eqnarray}

The coordinate operators $\hat{z'}$ and $\hat{\overline{z'}}$
share the same classical limit with $\hat{z}$ and
$\hat{\overline{z}}$ and differ only for terms of the order
$\frac{\sqrt{\theta}}{r}$ at large distance from the source.
However near the source we have $ \hat{z'} P_0 = P_0 \hat{z'} = 0$
while $ [ \hat{z}, P_0 ] \neq 0$.

To obtain the final result we have only to develop equation
(\ref{45}):

\begin{eqnarray}
Y & = & \frac{1}{2} ( Y_1 - i Y_2 ) = i [ X , \Lambda ] = i \left[
X, \hat{z'} \left( \frac{\tau_1 - i \tau_2}{2} \right) +
\hat{\overline{z'}} \left( \frac{\tau_1 + i \tau_2}{2} \right)
\right] \label{410}\end{eqnarray}

Using the Pauli matrices algebra

\beq \tau_3 ( \tau_1 \pm i \tau_2 ) = - ( \tau_1 \pm i \tau_2 )
\tau_3 = \pm ( \tau_1 \pm i \tau_2  ) \label{411}\eeq

we can simplify the result as

\begin{eqnarray}
Y & = & - \left( \frac{ \tau_1 - i \tau_2 }{2} \right) \
\sum^{\infty}_{n=0} \left[ \sqrt{n \left( n+1 + \frac{\mu}{2}
\right) } \
 | n+1 >< n+1 | \right. - \nonumber \\
& \ & -    \left. \sqrt{n \left( n+1 -\frac{\mu}{2} \right) }  \ |
n
>< n | \right] \nonumber \\
& \ & -  \left( \frac{ \tau_1 + i \tau_2 }{2} \right)
\sum^{\infty}_{n=0} \left[ \sqrt{ n \left( n + 2 -\frac{\mu}{2}
\right) } \
| n+2 >< n | \right. - \nonumber \\
& \ & -   \left. \sqrt{(n+1) \left( n + 1 +\frac{\mu}{2} \right)}
\ | n+2
>< n |\right] \label{412}\end{eqnarray}

\section{Classical limit}

To check the classical limit of the noncommutative solution ( eqs.
(\ref{323}) and (\ref{412}) ), we must recall the classical results
of ref. \cite{6}-\cite{7} in presence of a point source:

\begin{eqnarray}
e^a_\mu & = & \delta^a_\mu + \mu n^a n_\mu \ \ \ \ \ n_\mu =
\left( \epsilon_{ij} \frac{x^j}{r}, 0 \right) = \left(
\frac{y}{r} , - \frac{x}{r} , 0 \right) \nonumber \\
\omega_{\mu, ab} &  = & \epsilon_{ab3} \ \mu \ \frac{n_\mu}{r} \ \
\ \ \ \ \ \ \omega_i^a = - \mu \ \delta^a_3 \ \epsilon_{ij}
\frac{x^j}{r^2} \label{51}\end{eqnarray}

Working out the components we obtain

\begin{eqnarray}
\omega^3_z & = & \frac{1}{2} ( \omega^3_x - i \omega^3_y ) = -
\frac{i \mu}{2z} \nonumber \\
e^z_z & = & 1 - \frac{\mu}{2} \ \ \ \ e^{\overline{z}}_{z} =
\frac{\mu}{2} \frac{\overline{z}}{z} \label{52}\end{eqnarray}

The conical singularity becomes evident in the metric tensor built
from this form of the vierbein:

\begin{eqnarray}
g_{zz} & = & e^z_z e^{\overline{z}}_z = \frac{\mu}{2} \left( 1 -
\frac{\mu}{2} \right) \frac{\overline{z}}{z} \nonumber \\
g_{z \overline{z}} &  = & \frac{1}{2} ( e^z_z
e^{\overline{z}}_{\overline{z}} + e^{\overline{z}}_z
e^z_{\overline{z}} ) = \frac{1}{2} \left( 1 - \mu +
\frac{\mu^2}{2} \right) \nonumber \\
ds^2 & = & dz d \overline{z} + \frac{\mu}{2} \left( 1 -
\frac{\mu}{2} \right) r^2 {\left( \frac{ \overline{z} dz - z d
\overline{z} }{r^2} \right)}^2 = \nonumber \\
& = & dr^2 + {( 1 - \mu )}^2 r^2 d \phi^2 \label{53}\end{eqnarray}

At a noncommutative level it is simpler to compare the spin
connection given by eq. (\ref{319}) :

\beq X = \hat{p}_z + \frac{\omega^a_z}{2} \tau_a + \omega^0_z = i
\sum^{\infty}_{n=0} ( f(n) - \tau_3 g(n) ) \ | n+1 >< n |
\label{54}\eeq

Identifying the components we finally find:

\begin{eqnarray}
\omega^3_z & = & - 2  i \sum^{\infty}_{n=0} g(n) \ | n+1 >< n | =
- \frac{i}{\sqrt{ 2\theta}} \ \sum^{\infty}_{n=0} \left( \sqrt{
n+1 + \frac{\mu}{2} }
- \sqrt{ n+1-\frac{\mu}{2} } \right) \ | n+1 >< n | \nonumber \\
\omega^0_z & = & \frac{i}{2 \sqrt{ 2\theta}} \ \sum^{\infty}_{n=0}
\left( \sqrt{ n+1+\frac{\mu}{2} } + \sqrt{ n+1-\frac{\mu}{2} } - 2
\sqrt{ n+1 } \right) \ | n+1
>< n | \label{55}\end{eqnarray}

A first look shows that $\omega^3_z \sim O ( \mu ) $ while
$\omega^0_z \sim O ( \mu^2 ) $. However since the development
parameter is really $\frac{\mu}{n+1}$, every power in $\mu$
corresponds to a large distance behavior of the order
$\frac{\mu\sqrt{\theta}}{r}$. We therefore conclude that
$\omega^3_z$ has a nontrivial classical limit, while $\omega^0_z
\rightarrow 0$, as it should be.

Let us work out the first order contribution in $\mu$:

\beq \omega^3_z \approx - i \ \frac{\mu}{2\sqrt{2\theta}} \
\sum^{\infty}_{n=0} \ \frac{1}{\sqrt{ n+1 }} \ | n+1 >< n | = -
\frac{i\mu}{2 \hat{z}} \label{56}\eeq

and $\omega^3_z$ coincides with the operator corresponding to the
inverse of z, confirming that we have reached the right classical
limit, while $\omega^0_z$ contains only terms $O (
\frac{\mu\sqrt{\theta}}{r}) $ which vanish in the
$\theta\rightarrow 0$ limit.

The vierbein operator can be simplified by replacing

\beq f(n) \ \ \stackrel{\mbox{\small {$\theta \rightarrow 0$}
}}{\longrightarrow} \ \ \sqrt{\frac{ n+1 }{2\theta}} \ \ \ \ \ \
g(n) \ \ \stackrel{\mbox{\small {$\theta \rightarrow 0$}
}}{\longrightarrow} \ \   \frac{\mu}{4 \sqrt{2\theta} \sqrt{ n+1} }
\label{57}\eeq

and by approximating the factors $\sqrt{n} \approx \sqrt{ n+1 }$.
In this way we obtain the correct classical limit

\begin{eqnarray}
Y & \approx & \left( 1 - \frac{\mu}{2} \right) \left( \frac{ \tau_1
- i \tau_2 }{2} \right) + \frac{\mu}{4} \left( \frac{ \tau_1 + i
\tau_2 }{2} \right) \sum^{\infty}_{n=0} \left(
\sqrt{\frac{n+2}{n+1}} + \sqrt{\frac{n+1}{n+2}} \right) | n+2 >< n
| \nonumber \\
& + & O \left( \frac{ \sqrt{\theta} }{r} \right) = \left( 1 -
\frac{\mu}{2} \right) \left( \frac{ \tau_1 - i \tau_2 }{2} \right) +
\frac{\mu}{4} \left( \frac{ \tau_1 + i \tau_2 }{2} \right) \left(
\frac{1}{\hat{z}} \hat{\overline{z}} + \hat{\overline{z}}
\frac{1}{\hat{z}} \right) +  O \left( \frac{ \sqrt{\theta} }{r}
\right) \label{58}\end{eqnarray}

in perfect agreement with

\beq e^z_z = 1 - \frac{\mu}{2} \ \ \ \ \ e^{\overline{z}}_z =
\frac{\mu}{2} \left( \frac{\overline{z}}{z} \right) \label{59}\eeq

\section{The spinning case}

It is natural to generalize all these results to the case of a
massive and spinning source ( the $\mu \rightarrow 0$ is already
solved in ref. \cite{8} ). This is easily obtained by adding to the
massive solution an extra particular solution for the vierbein
corresponding to the torsion source.

Again the torsion source is defined through a representation of a
delta-function and the equations of motion to solve are

\begin{eqnarray} & &  [ \ X_\mu , X_\nu \ ] = \frac{i}{\theta} \ \epsilon_{\mu\nu 3} \
\left[ \ 1 - \frac{m}{4 \pi} P_0 \tau_3 \ \right] \nonumber \\
& & T_{\mu\nu} = [ X_\mu , Y_\nu ] - [ X_\nu , Y_\mu ] = - i \
\frac{s}{2 \pi \theta} \ \epsilon_{\mu\nu 3}  P_0 \tau_3 \ \ \ \ \ (
8 \pi G = 1 \ \ {\rm units } ) \label{61}\end{eqnarray}

As in the classical case, the solution is obtained by adding to the
vierbein operator (\ref{412}) an extra term $Y_\mu^S$:

\begin{eqnarray}
X_\mu & = & X_\mu ( \mu \neq 0 , s = 0 ) \nonumber \\
Y_\mu & = & Y_\mu ( \mu \neq 0 , s = 0 ) + Y^S_\mu
\label{62}\end{eqnarray}

In complex coordinates the torsion equation reads:

\beq [ \ X , \overline{Y}^S \ ] - [ \ \overline{X} , Y^S \ ] = \
\frac{s}{4\pi \theta} \ P_0 \tau_3 \label{63}\eeq

By introducing for the unknown operator $Y^S$ the same ansatz of
$X_\mu$:

\beq Y^S = i \sum^{\infty}_{n=0} \ ( h(n) - \tau_3 k(n) ) \ | n+1
>< n | \label{64}\eeq

The torsion equation produces the recursive relations for the
coefficients $h(n), k(n)$:

\begin{eqnarray}
& \ & f(n) h(n) + g(n) k(n) = 0 \nonumber \\
& \ & f(n) k(n) + g(n) h(n) = \frac{s}{8\pi\theta}
\label{65}\end{eqnarray}

which are solved by

\begin{eqnarray}
h(n) & = & \frac{s}{8\pi \sqrt{2\theta}} \left( \frac{1}{\sqrt{ n
+ 1 + \frac{\mu}{2} }} - \frac{1}{\sqrt{ n + 1 - \frac{\mu}{2} }} \right) \nonumber \\
k(n) & = & \frac{s}{8\pi \sqrt{2\theta}} \left( \frac{1}{\sqrt{ n
+ 1 + \frac{\mu}{2} }} + \frac{1}{\sqrt{ n + 1 - \frac{\mu}{2} }}
\right) \label{66}\end{eqnarray}

The reality conditions on $h(n)$ and $k(n)$ do not induce extra
constraints on $\mu$ other than the  limit $ \mu < 2 $. In the
massless limit we find agreement with the results of ref. \cite{8}:

\beq h(n) \rightarrow 0 \ \ \ \ k(n) \rightarrow \frac{s}{4\pi
\sqrt{ 2\theta }} \frac{1}{\sqrt{ n+1 }} \label{67}\eeq

\section{Measuring the deficit angle}

Since we have at disposition an explicit solution of
noncommutative gravity we can discuss its properties. For example,
we know that the conical metric can be revealed with an integral
of the distance around the source. The deficit angle is obtained
by choosing as a path a circle with the source in its center:

\beq \alpha = \frac{1}{r} \oint_C ds = \frac{1}{r} \int^{2\pi}_0 (
1 - \mu ) r d\phi = 2 \pi ( 1 - \mu ) \label{71}\eeq

Moreover this measure is not dependent on the distance from the
source.

What can we say at a noncommutative level ? Firstly to compute the
distance we need to build the metric given by the star product of
two vierbeins, calculated in eq. (\ref{412}). Our solution is
expressed only as an infinite series on the Laguerre polynomials. In
any case suppose that we have at disposition a formula in terms of
known functions for $ds$:

\begin{eqnarray} ds^2 & = & g_{\mu\nu}^{NC} dx^\mu dx^\nu \nonumber \\
g^{NC}_{\mu\nu} & = & \frac{1}{2} \{ ( e^a_\mu \tau_a + e^0_\mu)*
, ( e^b_\nu \tau_b + e^0_\nu ) \} \|_1 \label{72}
\end{eqnarray}

where the symbol $\|_1$ means that we need to extract the identity
part of this product, once it is expressed in the basis of the
$U(1,1)$ group.

Our particular formula (\ref{412}) for the vierbein generates a
correction $ O ( \frac{\sqrt{\theta}}{r} ) $ to the deficit angle:

\beq \alpha_{NC} = \frac{1}{r} \oint ds^{NC} = 2 \pi ( 1 - \mu ) +
O \left( \frac{\sqrt{\theta}}{r} \right) \label{73}\eeq

The problem we want to discuss is that it doesn't make sense to
look for a deterministic correction to the deficit angle in order
to characterize noncommutative gravity with respect to Einstein
gravity.

In a previous article \cite{5}, we have discussed how the metric (
and therefore the distance ) is not invariant under $U(1,1)$ gauge
group but it transform covariantly as

\begin{eqnarray}
G_{\mu\nu} & = & \frac{1}{2} \{ Y_\mu , Y_\nu \}\nonumber \\
 G_{\mu\nu} \ \ & \rightarrow  & \ \ U^{-1}
G_{\mu\nu} U \label{74}\end{eqnarray}

However we can still partially recover a continuity with classical
Einstein gravity by restricting the gauge group in order that at
spatial infinity the gauge transformations reduce to the identity.
The infinitesimal correction to $g_{\mu\nu}^{NC}$, due to the
application of this restricted gauge group, is

\beq \delta g_{\mu\nu}^{NC} \approx O \left(
\frac{\sqrt{\theta}}{r} \right) \label{75}\eeq

 i.e. the same order of the correction to the deficit angle
extrapolated by the explicit formula for the vierbein as in eq.
(\ref{412}).

Therefore the gauge group induces a statistical fluctuation of the
value of the deficit angle comparable with $ O(
\frac{\sqrt{\theta}}{r} )$.

Only at large distance from the source it is possible to recover
the typical determinism of general relativity. The same
observation applies to the geodesics around the source, whose
definition is dependent on the distance. Such an ambiguity of the
deficit angle automatically produces an ambiguity for the
scattering angle of a test particle in presence of a point-like
source, which can be avoided only for large impact parameter,
compared with the scale of the noncommutative parameter. This
ambiguity problem is an intriguing property of noncommutative
gravity which deserves a better understanding and it is probably
the distinguishing feature of such a theory.

Such considerations do not apply to the flat metric, which is
unaffected by the $U(1,1)$ gauge group. Therefore such a chaotic
effect here discussed requires the combination of a nontrivial
massive source with the noncommutativity of space-time.

\section{Conclusions}

Noncommutative gravity is a relatively new subject and in
literature rather few applications exist. The aim of this article
was to build a concrete and exactly solvable example, that helps
us in clarifying its physical meaning.

We have in fact succeeded in solving the noncommutative
gravitational field produced by a mass in ($2+1$) dimensions by
reformulating this problem in the operator formalism and making
use of the concept of a point-like source, an extended source
which only in $\theta \rightarrow 0$ limit produces a
delta-function singularity.  We show that it is possible to solve
the equations of motion with mass as unique source, responding to
ref. \cite{9}. The noncommutative gravitational field is then a
smooth deformation of the classical one, which can be recovered in
the $\frac{\sqrt{\theta}}{r} \rightarrow 0$ limit. Since the
source is extended, the noncommutative field is regular around the
source, while the classical one is singular.

Another question puzzles us and it could be the distinguishing
feature of such a theory, i.e. the gauge ambiguity of the distance
and consequently the geodesic field. Restricting gauge
transformations to be constant at spatial infinity we can recover
the classical determinism at large distance from the source.

Briefly speaking, noncommutative gravity contains two behaviors,
one which is deterministic at macroscopic level, giving coinciding
predictions with general relativity, and another one chaotic at
microscopic level.

\end{document}